\documentclass[sn-nature, pdflatex]{sn-jnl}

\usepackage{graphicx}%
\usepackage{multirow}%
\usepackage{amsmath,amssymb,amsfonts}%
\usepackage{amsthm}%
\usepackage{mathrsfs}%
\usepackage[title]{appendix}%
\usepackage{xcolor}%
\usepackage{textcomp}%
\usepackage{manyfoot}%
\usepackage{booktabs}%
\usepackage{algorithm}%
\usepackage{algorithmicx}%
\usepackage{algpseudocode}%
\usepackage{anyfontsize}
\usepackage{listings}%
\usepackage{lipsum}
\usepackage{cuted}
\usepackage{xstring}
\usepackage{siunitx}
\usepackage{xr-hyper}
\usepackage{hyperref}
\usepackage{natbib}
\usepackage{url}
\usepackage{circuitikz}
\usepackage[mathlines]{lineno}
\usepackage{tabularx}
\usepackage{placeins}
\DeclareSIUnit\torr{torr}

\DeclareSIUnit\sq{\ensuremath{\Box}}

\unnumbered{}

\newcommand{\upmu}[1]{\si{\micro\ampere}}

\begin{document}

\title[Superconducting Nanowire Memory Array]{Scalable Superconducting Nanowire Memory Array with Row-Column Addressing}
\subtitle{\today}

\author*[1]{\fnm{Owen} \sur{Medeiros}}\email{omedeiro@mit.edu}

\author[1]{\fnm{Matteo} \sur{Castellani}}
\author[1]{\fnm{Valentin} \sur{Karam}}
\author[1]{\fnm{Reed} \sur{Foster}}
\author[1]{\fnm{Alejandro} \sur{Simon}}
\author[1]{\fnm{Francesca} \sur{Incalza}}
\author[1]{\fnm{Brenden} \sur{Butters}}
\author[1, 2]{\fnm{Marco} \sur{Colangelo}}

\author*[1]{\fnm{Karl K.} \sur{Berggren}}\email{berggren@mit.edu}

\affil[1]{\orgdiv{Department of Electrical Engineering and Computer Science}, \orgname{Massachusetts Institute of Technology}, \orgaddress{\street{77 Massachusetts Ave.}, \city{Cambridge}, \postcode{02139}, \state{Massachusetts}, \country{USA}}}

\affil[2]{\orgdiv{Department of Electrical and Computer Engineering}, \orgname{Northeastern University}, \orgaddress{\street{360 Huntington Ave}, \city{Boston}, \postcode{02115}, \state{Massachusetts}, \country{USA}}}

\abstract{
    Developing ultra-low-energy superconducting computing and fault-tolerant quantum computing will require scalable superconducting memory.
    While conventional superconducting logic-based memory cells have facilitated early demonstrations, their large footprint poses a significant barrier to scaling.
    Nanowire-based superconducting memory cells offer a compact alternative, but high error rates have hindered their integration into large arrays.
    In this work, we present a superconducting nanowire memory array designed for scalable row-column operation, achieving a functional density of 2.6$\,$Mb/cm$^{2}$.
    The array operates at $\num{1.3}\,\si{\kelvin}$, where we implement and characterize multi-flux quanta state storage and destructive readout.
    By optimizing write and read pulse sequences, we minimize bit errors while maximizing operational margins in a $\num{4}\times\num{4}$ array.
    Circuit-level simulations further elucidate the memory cell's dynamics, providing insight into performance limits and stability under varying pulse amplitudes.
    We experimentally demonstrate stable memory operation with a minimum bit error rate of 10$^{-5}$.
    These results demonstrate feasibility for larger superconducting nanowire memory arrays, with potential applications in superconducting electronic architectures.
}


\maketitle

\section{Main}\label{sec1}

The advancement of superconducting computing and fault-tolerant quantum systems depends on developing scalable, energy-efficient cryogenic memory architectures~\cite{alam2023cryogenic}.
While superconducting logic circuits have progressed significantly, the lack of dense and reliable memory has remained a critical bottleneck.
Existing superconducting memories, primarily based on Josephson junctions (JJs), face fundamental challenges in scaling due to their large unit cell footprints~\cite{holmes2021cryogenic, herr2023superconducting}.
These limitations have motivated the exploration of alternative device architectures and materials that can reduce bit-cell area while maintaining low power operation and non-volatility~\cite{golod2023word}.

Superconducting loops have long been explored for cryogenic memory applications due to their ability to store information in the form of persistent circulating currents, offering zero static power dissipation and nonvolatile storage while in the superconducting state~\cite{anacker1969potential, zappe1974single, chen1993nondestructive,karamuftuoglu2024superconductor}.
Traditional JJ-based loop memories, including vortex-transition (VT) cells~\cite{semenov2019very}, have demonstrated functional densities of $1\,\si{\mega\bit\per\centi\meter\squared}$.
However, the geometric inductance required to accommodate a single flux quantum, $\Phi_0 = LI$, constrains further scaling, since reducing the loop area demands a higher drive current to induce a phase transition~\cite{golod2023word}.
This trade-off between cell size and power consumption has increased interest in alternative memory designs that bypass these geometric constraints~\cite{volk2023addressable}.

Scaling superconducting memory to smaller footprints has driven the exploration of alternative loop-based designs beyond conventional Josephson junctions.
Abrikosov vortex RAM (AVRAM) achieves the smallest reported memory cell, with a bit area of $1\,\si{\micro\meter\squared}$ using focused ion beam junctions and vortex traps~\cite{golod2023word}.
However, this approach has not yet been demonstrated in large arrays and may face integration challenges due to its reliance on non-standard fabrication processes.

Superconducting nanowire memory (SNM or nMem) offers a scalable alternative by exploiting kinetic inductance—an intrinsic material property—rather than geometric inductance, which depends on physical loop dimensions.~\cite{murphy2017nanoscale, zhao2018compact, mccaughan2018kinetic, ilin2021supercurrent}.
This approach allows smaller memory cells since kinetic inductance ($L_k$) can be much larger than geometric inductance ($L_g$) without increasing the physical footprint.
A prior demonstration of an $8\,\si{bit}$ SNM array achieved a functional density exceeding $\num{1}\,\si{\mega\bit\per\centi\meter\squared}$~\cite{butters2021scalable}, validating the viability of planar row–column addressing architectures for nanowire-based memory.

While these early results established feasibility, practical scalability remained limited by high error rates during word-line activation.
These errors arose from a disproportionate increase in cell inductance, which reduced the separation between readout signal levels and increased bit errors~\cite{butters2021scalable}.

To address this challenge, we introduce a variable kinetic inductor that preserves the inductance asymmetry between memory cell branches at elevated operating temperatures, ensuring stable signal margins during array operation.
This innovation reduced the bit error rate from $10^{-3}$ to $10^{-5}$ and enabled reliable operation across a full 16-cell array with 15 out of 16 cells outperforming the previous state-of-the-art.

In this work, we fabricated and characterized a $16$-bit SNM array using a planar row-column architecture with a functional density of $2.6\,\si{\mega\bit\per\centi\meter\squared}$.
Although our array density only moderately surpasses the state-of-the-art ($1\,\si{\mega\bit\per\centi\meter\squared}$)~\cite{semenov2019very}, it reflects an unoptimized design.
Higher integration densities may be achievable through targeted layout strategies (e.g., reducing cell area~\cite{buzzi2023nanocryotron} and closer cell spacing), use of materials with higher kinetic inductance, and fabrication improvements toward \SI{50}{\nm} critical dimensions.
The array operated at a base temperature of $1.3\,\si{\kelvin}$ and supported destructive read and write operations using short voltage pulses.
We optimized pulse parameters to minimize bit error rate while maximizing threshold margins.
We achieved a minimum bit error rate (BER) of $1 \times 10^{-5}$ at $1\,\si{MHz}$ and demonstrated robust memory retention across hold times from $2\,\si{\micro\second}$ to $20\,\si{\second}$.
The measured performance confirms the viability of SNM arrays for scalable cryogenic memory applications.

\section{Superconducting Nanowire Memory Array}\label{sec:MemoryArray}

\begin{figure*}[htbp] \centering \includegraphics[width=\textwidth]{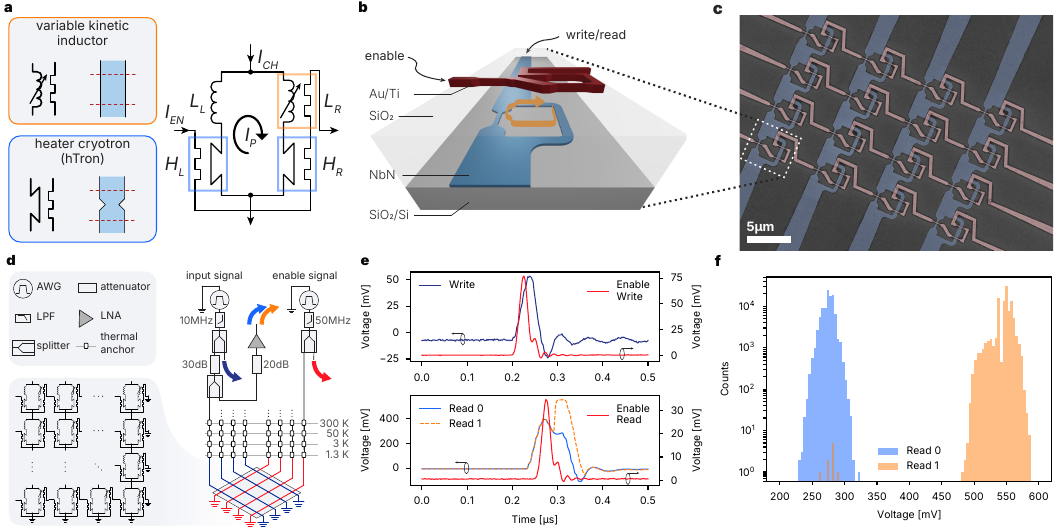} \caption{ \textbf{Superconducting nanowire memory array.}
        \textbf{a}, Schematic of a memory unit cell composed of a superconducting loop incorporating two hTrons and a variable kinetic inductor.
        The cell is biased via two current ports labeled $I_{\text{EN}}$ and $I_{\text{CH}}$, corresponding to the enable and channel (write/read) currents, respectively.
        The superconducting channel is shown in blue, and the heater location is indicated by red dashed lines.
        An additional constriction in the hTron ensures switching occurs preferentially at the heater site.
        \textbf{b}, Three-dimensional rendering of the memory cell stack.
        A NbN film, patterned on a thermal oxide wafer, forms the superconducting columns.
        A gold layer defines the normal-metal rows and is galvanically isolated from the NbN by a $\num{100}\,\si{\nano\meter}$ silicon dioxide layer.
        \textbf{c}, False-colored scanning electron micrograph of a $\num{16}\,\si{\bit}$ memory array, highlighting the superconducting columns (blue) and normal-metal rows (red).
        \textbf{d}, Schematic of the experimental setup.
        Bit-line (input) and word-line (enable) signals are delivered to the array, which is mounted on a PCB and thermalized to $\num{1.3}\,\si{\kelvin}$ inside a closed-cycle cryostat.
        Abbreviations: AWG, Arbitrary Waveform Generator; LPF, Low Pass Filter; LNA, Low Noise Amplifier.
        \textbf{e}, Measured voltage traces during a write operation.
        Signals associated with the superconducting column are plotted with respect to the left axes; those associated with the normal-metal row are referenced to the right axes.
        The write-pulse (blue) encodes logical states via pulse polarity (positive for 1, negative for 0).
        The enable-pulse (red), applied concurrently, modulates the critical current of the cell.
        Measured pulse parameters are listed in Table~\ref{tab:table1}.
        Typical read voltage traces for both logical states are shown in the bottom panel (0 in light blue, 1 in orange).
        \textbf{f}, Histograms of read voltages for both logical states over $\num{200e3}$ measurements.
        Nine W1R0 (write 1, read 0) errors were observed, with zero W0R1 (write 0, read 1) errors.
        The broader distribution for read 1 arises from variations in hot spot size, whereas the narrower read zero distribution reflects the absence of a hot spot and depends on the read-current amplitude.}\label{fig:figure1}
\end{figure*}

The basic operation relies on storing information as the direction of current flow in a superconducting loop.
To write information, we apply a current pulse that selectively switches one of two superconducting constrictions in the loop branches, establishing a persistent current that circulates clockwise or counter-clockwise to represent binary states 0 and 1.
Changing the sign of the write pulse reverses the direction of the induced persistent current.
Control of the persistent current is achieved using temperature-dependent superconducting switches or heater-cryotrons (hTrons)~\cite{zhao2018compact, baghdadi2020multilayered,karam2024parameter, wang2025attojoule}.
The memory cell contains two hTrons ($H_L$ and $H_R$), one for each branch of the superconducting loop, shown in Figure~\ref{fig:figure1}a.

To ensure robust and deterministic switching, we chose the write-pulse amplitude so that the left hTron ($H_L$) transitions to the normal state while the right hTron ($H_R$) remains superconducting.
We achieved this selective switching by engineering asymmetries in both critical currents and loop inductances: we designed $H_L$ with a lower critical current than $H_R$ (constriction widths of \SI{100}{\nm} and \SI{300}{\nm}, respectively) and set the loop inductance such that the inductance of the left branch is less than the inductance of the right branch ($L_{\text{L}} \ll L_{\text{R}}$), where the branch inductance is dominated by the kinetic inductance.
As a result, a larger fraction of the write-current flows through the left branch, causing $H_L$ to switch before $H_R$.
After we remove the pulse, the resulting circulating current becomes trapped in the loop, preserving the logical state.

Although the loop asymmetry enables deterministic switching in an isolated cell, this design does not inherently support scalable array operation.
The architecture includes an additional control mechanism to enable multiple cells within a column: an enable current passing through a~\SI{100}{\nm} wide gold trace that selectively heats the targeted cell.
This local heating lowers the critical current of the superconducting channel, facilitating switching, but also increases its inductance.
Compensating for this thermally induced inductance shift is central to ensuring scalable memory operation.

The fundamental switching element is the hTron~\cite{zhao2018compact, baghdadi2020multilayered, alam2023reconfigurable,wang2025attojoule}, a superconducting electrothermal device.
The hTron consists of a superconducting channel that is locally driven normal via Joule heating from an adjacent normal-metal heater.
Its switching threshold is governed by the heater (or enable) current, $I_{\text{enable}}$, which modulates the channel's critical current, as characterized by Karam et al.~\cite{karam2024parameter}.\footnote{The hTron can also function as a standalone memory element, with the stored state encoded in the impedance of the superconducting channel~\cite{wang2025attojoule}.}

In addition to the hTron, each memory cell includes a variable kinetic inductor that remains superconducting but exhibits a temperature-dependent kinetic inductance, $L_{\text{K}}(T)$.
Structurally, it is identical to the hTron aside from a constriction that locally reduces the critical current.
When the enable-current is applied, the resulting local heating increases the inductance of the right branch, maintaining the loop asymmetry condition across thermal states: $L_{\text{L}}(T_{\text{sub}}) \ll L_{\text{R}}(T_{\text{sub}})$ at the substrate temperature, and $L_{\text{L}}(T_{\text{enable}}) \ll L_{\text{R}}(T_{\text{enable}})$ when the enable-current is active.
Where $T_{\text{sub}}$ is the substrate temperature and $T_{\text{enable}}$ is the channel temperature during enable operation.
This dynamic inductance compensation, ensuring the left branch inductance is always less than the right branch inductance, is critical for reliable operation in an array configuration~\cite{butters2022digital}.

The memory array was implemented by tiling the SNM unit cell in a row–column architecture, enabling scalable addressing.
Each unit cell consists of two hTrons and a variable kinetic inductor, as shown in Figure~\ref{fig:figure1}a.
The circuit elements were patterned from a $\num{23}\,\si{\nano\meter}$-thick niobium nitride (NbN) superconducting thin film deposited on a thermal oxide wafer.
A normal-metal enable line, galvanically isolated from the superconducting layer by a $\num{100}\,\si{\nano\meter}$ silicon dioxide layer deposited via plasma-enhanced chemical vapor deposition (PECVD), provides local temperature control during write and read operations.
Figure~\ref{fig:figure1}b shows a three-dimensional rendering of the memory stack-up, and Figure~\ref{fig:figure1}c displays a false-colored scanning electron micrograph of the fabricated $\num{16}\,\si{\bit}$ array.

Thermal crosstalk between adjacent cells is minimal due to the highly localized nature of hTron heating.
The array density used here was chosen to ensure negligible thermal interaction between cells; denser packing is possible with additional design considerations.

The array was wirebonded to a printed circuit board (PCB) and thermally anchored to a $\num{1.3}\,\si{\kelvin}$ stage in a closed-cycle cryostat (Figure~\ref{fig:figure1}d).
Room-temperature electronics supplied the input (bit line) and enable (word line) signals to the device.

Figure~\ref{fig:figure1}e presents measured voltage traces during a write operation, corresponding to the signal paths (colored arrows) indicated in panel \textbf{d}.
The write-pulse polarity determines the direction of the stored persistent current, while a subsequent, unipolar read-pulse retrieves the stored logical state.
Each trace has a time average of over 500 consecutive acquisitions to improve the signal-to-noise ratio.

The memory does not have a controlled initial state and may contain trapped flux.
However, when the initial state matters, the system can clear flux by driving the loop into the normal state.

The bottom panel of Figure~\ref{fig:figure1}e shows typical readout voltages for logical 0 and 1 states.
Histograms of the read voltages across $\num{200e3}$ repeated measurements are shown in Figure~\ref{fig:figure1}f.
We observed nine W1R0 (write 1, read 0) errors and no W0R1 (write 0, read 1) errors.
The shape of the read one distribution is broader due to the additional variability introduced by the hot spot.

\FloatBarrier{}
\section{Characterization of Memory Cell Performance}\label{sec:Characterization}

We characterized memory cells by measuring the BER, defined as the fraction of operations where the read state differed from the written state, under varying read and write bias conditions.
To measure the BER, we wrote a known logical state (0 or 1), waited a set delay, and then read the state by analyzing the amplitude of the read voltage pulse, visible near \SI{310}{\ns} in the blue and orange traces of Figure~\ref{fig:figure1}e.
This process was repeated across many cycles, and readout voltages were classified according to the expected logical state.
A mismatch between the expected and measured state was counted as an error, and the BER was calculated as the ratio of errors to total trials.

Because logical states are arbitrarily assigned, a BER near 0 indicates correct operation with the chosen assignment, while a BER near 1 corresponds to error-free operation under the opposite assignment.
A BER of 0.5 reflects random behavior with indistinguishable logical states.

Figure~\ref{fig:figure1}f demonstrates a representative bimodal distribution of read voltage amplitudes corresponding to the logical states.
BER is computed as previously defined, and sweeping the read-current amplitude reveals an optimal bias point where the BER is minimized.

The direction of the persistent current—either nominal or inverted—is determined by the relative timing at which each hTron returns (i.e. retraps) to the superconducting state during the write operation. Initially, if the write-current amplitude exceeds only the critical current of the left hTron, the left channel becomes resistive first. After the write-pulse ends, the left hTron retraps to the superconducting state while the right hTron remains superconducting throughout, resulting in a nominal persistent current direction.

In this scenario, as the write pulse decays, the left hTron typically cools and returns to the superconducting state before the right hTron because its hotspot is smaller.
This leads to earlier retrapping and establishes a persistent current in the inverted direction.
However, if both branches switch and retrap nearly simultaneously, or if switching oscillations occur, the final state may still be non-inverting depending on the exact retrapping sequence.
\footnote{This behavior is confirmed by simulation results and explicitly illustrated in Fig.~\ref{fig:figure3}, which details the current dynamics and switching sequence underlying both nominal and inverted operation modes}.

\begin{figure*}[htbp]
    \centering
    \includegraphics[width=\textwidth]{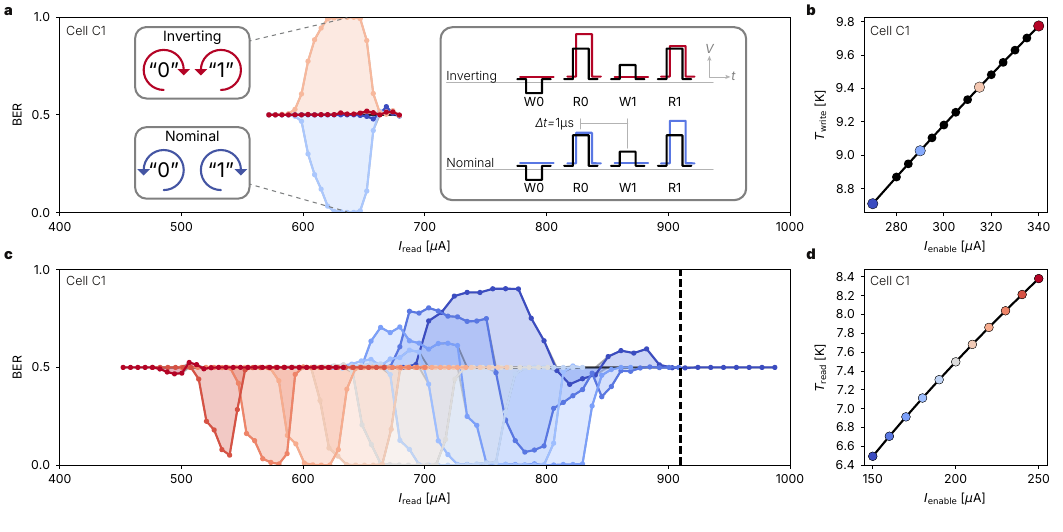}
    \caption{
        \textbf{Readout Fidelity and Operational Modes of a Temperature-Enabled Superconducting Memory Cell.}
        \textbf{a}, Measured BER as a function of read current amplitude at four different write temperatures.
        Each BER trace contains 31 points, with 1000 read measurements per point.
        A BER approaching 1 corresponds to a deterministic inverting mode, distinct from random errors.
        The four operational modes are: no switch (blue), nominal (light blue), inverting (light red), and switched (red).
        Insets with leaders highlight data points where the BER approaches 0 and 1, and indicate the direction of the persistent current for each state.
        The large inset shows a sketch of voltage versus time describing the write and read inputs used for the BER measurement.
        Nominal and inverting output traces are sketched in light blue and light red, respectively.
        \textbf{b}, Channel temperature during the write operation as a function of enable current, calculated using equation~\ref{eq:channel_temperature}. Colored markers correspond to the read-current sweeps in \textbf{a}.
        \textbf{c}, Measured BER as a function of read current amplitude for a sweep of read temperatures.
        The sweep is performed at temperatures between $\num{6.5}\,\si{\kelvin}$ (blue) and $\num{8.4}\,\si{\kelvin}$ (red), set by the enable-current.
        The write-current is held constant at $\num{30}\,\si{\micro\ampere}$.
        The vertical dashed line indicates the critical current of the memory cell without the enable current.
        \textbf{d}, Channel temperature during the read operation as a function of enable current.
        Colored markers correspond to the read temperature sweeps in \textbf{c}.
    }\label{fig:figure2}
\end{figure*}

\begin{figure*}[htbp]
    \centering
    \includegraphics[width=0.9\textwidth]{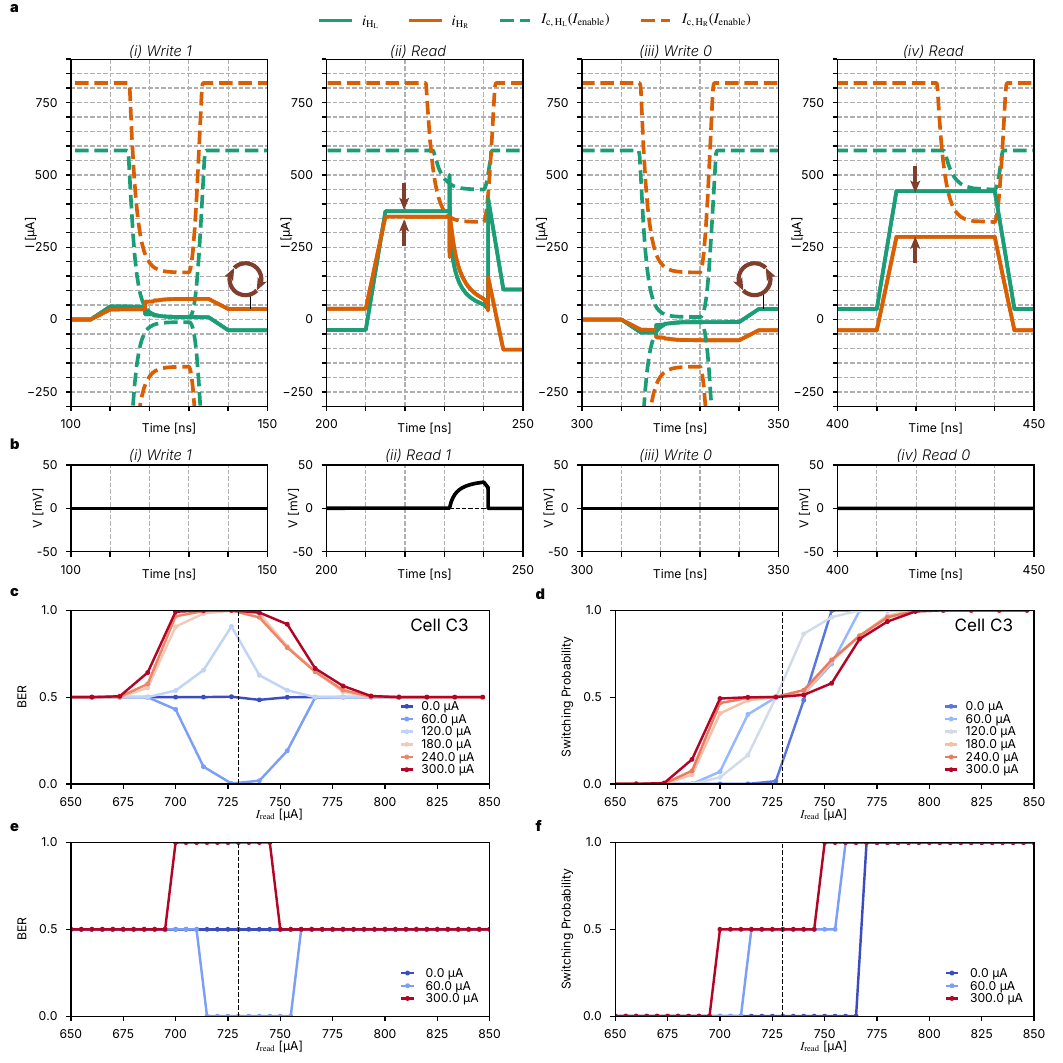}
    \caption{\textbf{Simulated and measured operating bounds of the memory cell.}
        \textbf{a}, Simulated current amplitudes through the left ($H_{L}$, green) and right ($H_{R}$, orange) hTrons during memory operations.
        Dashed lines indicate the corresponding critical currents, $I_{c, H_L}(I_{\text{enable}})$ and $I_{c, H_R}(I_{\text{enable}})$.
        Panels i-iv correspond to write-one (W1), write-zero (W0), read-one (R1), and read-zero (R0) operations, respectively.
        The hTron critical current, modulated by the heater ($I_{\text{enable}}$), is plotted as both a positive and negative value for reference to the write-current amplitude.
        After the write operation, we note the direction of the persistent current with rotating arrows.
        Before the read operation, arrows show the difference in branch currents due to the persistent current.
        \textbf{b}, Simulated output voltage across the memory cell during each operation in Fig.~\textbf{a}.
        No voltage is present during write operations (i, iii), where only $H_{L}$ transitions to the normal state.
        A voltage appears during the read-one operation (ii), while the read-zero operation (iv) yields no output, demonstrating nominal readout behavior.
        \textbf{c}, Measured BER as a function of read current amplitude for various write-current amplitudes.
        The cell does not operate when $I_{W} = 0$, shows nominal behavior for $I_{W} < \num{120}\,\si{\micro\ampere}$, and exhibits inverted operation for $I_{W} > \num{120}\,\si{\micro\ampere}$.
        \textbf{d}, Measured switching probability (i.e., voltage detection) during the read operation of cell C3 as a function of read current amplitude.
        At low read currents, the memory cell rarely switches, while at high currents, switching is nearly deterministic.
        \textbf{e}, Simulated BER for read current sweeps at three representative write-currents: 0, 60, and $\num{300}\,\si{\micro\ampere}$.
        The trends closely match those observed in panel \textbf{c}, with nominal and inverted regimes depending on write amplitude.
        \textbf{f}, Simulated switching probability versus read current amplitude for the same three write-currents as in \textbf{e}.
        The behavior mirrors that in panel \textbf{d}, with a sharp transition from low to high switching probability.
        The black dashed line in \textbf{c}-\textbf{f} marks the read-current amplitude used in the simulations shown in panels \textbf{a} and \textbf{b}.
    }\label{fig:figure3}
\end{figure*}

At the lowest write temperature (marked blue in Fig.~\ref{fig:figure2}a,b), the critical current of each hTron was greater than the write current, and therefore no persistent current was stored during the write operation, resulting in a BER of 0.5 for all read current amplitudes.
At the highest write temperature (red markers), both hTron critical currents are exceeded, similarly resulting in no persistent current.

The two intermediate write temperatures (light blue and light red in Fig.~\ref{fig:figure2}a,b) allowed for the channel to be programmed to either a nominal or inverting state, respectively.
The direction and assigned state convention are shown in the inset of Fig.~\ref{fig:figure2}a with leaders connected to the data points where the BER approaches zero and one.
Both extremes, approximately \SI{290}{\micro\ampere} and \SI{320}{\micro\ampere}, thus mark operational limits, while intermediate temperatures allow deterministic programming.
At each read current amplitude point, the BER was calculated from $\num{1e3}$ read measurements, ensuring sufficient statistical reliability.

Each measurement in the BER sweep consists of a sequence of write and read operations, executed with a typical period of $\num{1}\,\si{\micro\second}$, as illustrated in the larger inset of Fig.~\ref{fig:figure2}a.
The channel input signal is shown in black, with each operation labeled below.
The output signal is shown in light blue for nominal operation, where a write-one (W1) produces a voltage pulse, and in light red for the inverting case, where a write-zero (W0) produces a voltage pulse.
The W0 and W1 pulses had equal amplitudes, differing only in sign, while the read-zero (R0) and read-one (R1) pulses were identical.

Additional pulse sequences were tested to verify that memory cell reliability was independent of input sequence variations.
These tests included extra read pulses simulating interactions with other cells in the same column and additional enable pulses to emulate row-access scenarios.
Sequences involving sustained activation of the enable line were also investigated to detect potential thermal latch-up effects.
None of these variations resulted in an increased BER, confirming robustness against sequence-dependent errors.
Further investigation into additional sequence patterns, specifically for the addressing of multiple cells (write to a cell, perform operations on the rest of the array, read from the cell), remains important to conclusively rule out sequence-dependent errors.

Figure~\ref{fig:figure2}b presents the calculated channel temperature during write operations as a function of the applied enable-current, derived using the thermal model described by Equation~\ref{eq:channel_temperature}.
Modulating the enable current effectively tunes the memory cell's temperature, thus controlling the critical currents of the hTrons and enabling the programming of distinct memory states.
The resulting temperature range is between \num{72} and \SI{76}{\percent} of the critical temperature, $T_c$, of the NbN film, which is approximately $\num{12.5}\,\si{\kelvin}$, indicating the critical current of the selected cell is significantly reduced (see Fig.~\ref{fig:figure2}b).

Figure~\ref{fig:figure2}c displays BER results from read current sweeps conducted at varying read temperatures, set by adjusting the enable current.
In contrast to the write-temperature results (Fig.~\ref{fig:figure2}a), varying the read temperature shifted the optimal read current region.
The optimal read temperature, approximately $\num{7.5}\,\si{\kelvin}$, provided a favorable balance: sufficiently elevating the critical current difference between the target and neighboring cells (thus minimizing BER), while remaining below the threshold for thermal latch-up, ensuring robust and stable device operation.
The vertical dashed line denotes the memory cell's critical current without the enable current applied, providing a reference point to contextualize the broad operational range of the device.

Finally, Figure~\ref{fig:figure2}d presents the calculated channel temperature during the read operation as a function of the enable current.
This temperature was calculated using Equation~\ref{eq:channel_temperature}.

Figure~\ref{fig:figure3} illustrates how circuit simulations reproduce the observed memory behavior.
Panel (a) shows simulated hTron currents during the four basic operations, capturing the selective switching and persistent current storage.
The resulting voltage responses align with measured BER sweeps (c–d), and are well-matched by simulated sweeps (e–f), confirming the model's consistency with experimental operation.

In Figure~\ref{fig:figure3}a the loop begins with zero current and the hTron critical currents at the maximum value.
Panel (i) shows the write-one operation, consisting of an applied write bias ($I_{\text{write}}=i_{H_L}+i_{H_R}$) and an applied enable bias ($I_{\text{enable}}$).
The applied write bias splits inductively between the two channels, yielding $i_{H_L} > i_{H_R}$.
At $\num{120}\,\si{\nano\second}$, the left hTron switches when $i_{H_L} > I_{c,H_L}(I_{\text{enable}})$, while the right hTron remains superconducting $i_{H_R} < I_{c,H_R}(I_{\text{enable}})$.
Only left branch switches because the left hTron has a lower critical current and receives a larger fraction of the write current due to the inductive asymmetry.
Current redistribution and removal of the write bias establishes a clockwise persistent current, $I_{\text{P}} \approx I_{\text{write}}$.

This persistent current remains in the loop at the start of the read-one operation (ii).
The applied read bias adds to $i_{H_R}$, pushing it above threshold ($i_{H_R} > I_{c,H_R}$), while $i_{H_L} < I_{c,H_L}$ due to the opposing persistent current.
Current then redistributes, causing the left hTron to switch and generate a voltage pulse (Fig.~\ref{fig:figure3}b).
This destructive readout resets the cell; it must be rewritten after each read.
The presence or absence of a voltage pulse during readout corresponds to logical states `1` and `0`, respectively.

To return the cell to the zero state, the write-zero operation (iii) follows the same procedure as (i), but with an inverted write bias.
The left hTron switches ($i_{H_L} > I_{c,H_L}$), while the right hTron remains superconducting, and the resulting persistent current is counter-clockwise, $I_{\text{P}} \approx -I_{\text{write}}$.

Readout in (iv) follows the same sequence as (ii), but with an inverted persistent current present.
This reversal modifies the readout conditions: the persistent current now adds to $i_{H_L}$ and subtracts from $i_{H_R}$, opposite to the read-one case.
As indicated by the arrows in panel~(iv), $i_{H_L}$ increases but remains below $I_{c,H_L}$, while $i_{H_R}$ is further suppressed and stays below $I_{c,H_R}$.
Neither hTron switches, and no voltage is produced, as shown in Fig.~\ref{fig:figure3}b.
The opposite outcomes of panels~(ii) and~(iv) reflect the sign of the persistent current and enable binary memory operation.

The read current used in panels~a and~b is indicated by a dashed line in plots~c–f.

Although the design targets symmetric heating, the precise temperatures and resulting critical currents of the hTrons are unknown.
To account for this and fabrication-related variations, the temperature of each hTron was adjusted independently.

Figure~\ref{fig:figure3}c plots a measured dataset similar to Fig.~\ref{fig:figure2}a, but at a fixed write temperature (enable-write current) and varying write-current amplitudes.
Sweeping the write-current amplitude changes the magnitude of the persistent current in the memory cell.
At low amplitudes, both hTrons remain superconducting and no current is written to the loop, yielding a BER of $\num{0.5}$ for all read currents.
When the write-current amplitude exceeds the left hTron critical current during the write operation, the current is diverted to the right hTron, writing a nominal persistent current.
Continued increase of the write current leads to the same behavior until the remaining current after retrapping in the left hTron exceeds the right hTron’s critical current, i.e., when $I_{\text{write}} - I_{r,H_L} > I_{c,H_R}$.

In Figure~\ref{fig:figure3}c, the transition between nominal and inverted operation occurs at approximately $\num{120}\,\si{\micro\ampere}$.
The inverted persistent current is a result of the left hTron returning to the superconducting state before the right hTron.

Figure~\ref{fig:figure3}d plots the measured switching probability as a function of read current amplitude.
At low read currents, the memory cell rarely switches; at high currents, switching occurs with near-unity probability.

Figure~\ref{fig:figure3}e presents simulated BER for read-current sweeps at three representative write amplitudes: $\num{0}$, $\num{60}$, and $\num{300}\,\si{\micro\ampere}$.
The trends capture the same nominal and inverted regimes as in panel~c.
Quantitative agreement requires further refinement of the simulation model.

Figure~\ref{fig:figure3}f shows the simulated switching probability under the same conditions, which mirrors the behavior in panel~d.
The probability is defined as the ratio of read operations that produce a voltage pulse to the total number of read operations.

\FloatBarrier{}
\section{Array Operating Limits}\label{sec:OperatingLimits}

For a superconducting memory array to function reliably, precise input parameter selection is essential to prevent unintended state changes in neighboring cells.
Column inputs must adhere to the constraint that the sum of read and persistent currents remains below the lowest critical current among unselected memory cells, described as $(I_{\text{read}}+I_{\text{P}})<I_{\text{c}}(T_{\text{sub}})$.
Similarly, row inputs are limited by conditions ensuring the critical current of the selected memory cell, reduced by the enable-current, remains greater than the persistent current ($I_{\text{c}}(T_{\text{enable}})>I_{\text{P}}$), while still being lower than that of unselected cells ($I_{\text{c}}(T_{\text{enable}})<I_{\text{c}}(T_{\text{sub}})$).

\begin{figure*}[htbp]
    \centering
    \includegraphics[width=0.9\textwidth]{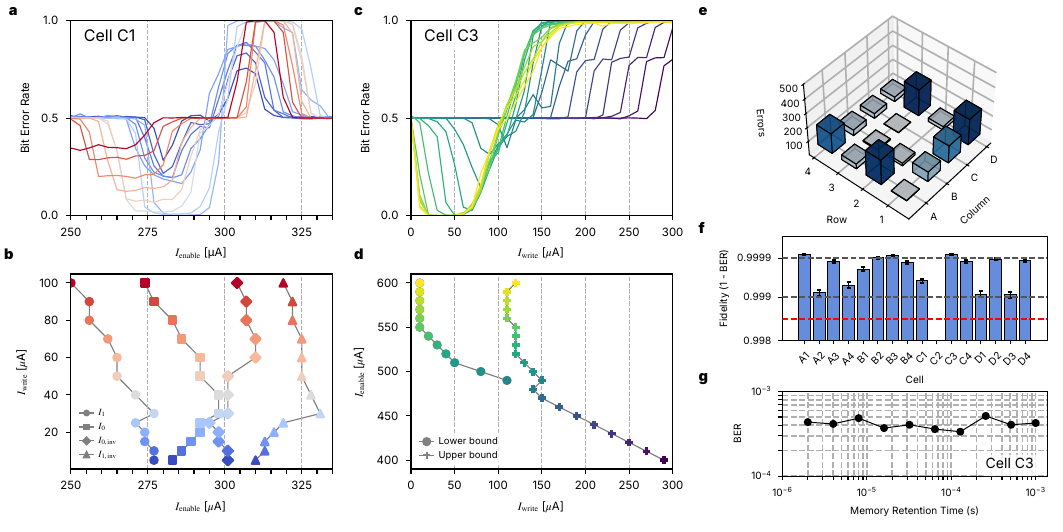}
    \caption{
        \textbf{Operating limits and parameter sensitivity across a memory array.}
        \textbf{a}, Measured BER of cell C1 as a function of enable current, evaluated across several fixed write-current amplitudes ranging from $\num{5}\,\si{\micro\ampere}$ (blue) to $\num{100}\,\si{\micro\ampere}$ (red).
        At low write currents, two distinct peaks appear, corresponding to nominal and inverted switching.
        As the write current increases, the peaks broaden and fidelity improves.
        Beyond $400~\si{\micro\ampere}$, nominal fidelity degrades as the optimal operating region shifts to lower enable currents and becomes narrower.
        \textbf{b}, Enable current bounds extracted from the data in \textbf{a}, representing the operational window for each write-current.
        Bounds are defined as the range over which the BER deviates from 0.5 by more than 5\%.
        \textbf{c}, BER of cell C3 as a function of write-current, for a set of enable current amplitudes ranging from $400$ to $600~\si{\micro\ampere}$ in $10~\si{\micro\ampere}$ steps.
        Higher enable currents raise the channel temperature during the write operation.
        \textbf{d}, write-current bounds extracted from the data in \textbf{c}, using the same $\pm5\%$ BER deviation criterion.
        The lower bound corresponds to the minimum write current required to switch the left hTron, while the upper bound marks the regime in which both hTrons exceed their critical currents.
        \textbf{e}, Number of errors during 200k measurement operations for each cell in a $\num{4}\times\num{4}$ array, using a $2~\si{\micro\second}$ hold time between write and read operations.
        Most cells exhibit BERs below $10^{-3}$, except C2, which shows elevated error likely due to a reduced critical current in its neighboring memory cell, C3.
        No systematic spatial variation is observed across the array.
        \textbf{f}, Fidelity (1-BER) for each cell in the array, showing all cells achieve greater than 99.9\% fidelity, except C2 (94\% fidelity).
        The dashed red line indicates the best fidelity achieved in previous work~\cite{butters2021scalable}.
        \textbf{g}, BER as a function of the delay between write and read operations.
        A single bias point was used for all measurements to emphasize the temporal stability of the memory cell rather than minimize BER at each time point.
        Each data point represents the average of 200k trials and the errorbars expected from binomial statistics are smaller than the markers.}\label{fig:figure4}
\end{figure*}

To establish these operating constraints, we extensively characterized the bit-error rate (BER), as summarized in Figure~\ref{fig:figure4}.
Figure~\ref{fig:figure4}a illustrates the measured BER for cell C1 as a function of the enable-write current ($I_{\text{enable}}$), with distinct traces corresponding to write-currents ($I_{\text{write}}$) ranging from $\num{5}\,\si{\micro\ampere}$ (blue) to $\num{100}\,\si{\micro\ampere}$ (red).
Increasing $I_{\text{enable}}$ gradually lowers the critical current of the memory cell, initially causing the left hTron to switch and redirecting current into the right hTron branch.
If the redirected current also exceeds the right hTron's critical current, both branches become resistive. As the bias pulse decays, the left hTron typically cools and returns to the superconducting state first due to its smaller hotspot.
This retrapping order reverses the direction of circulating current in the loop, effectively programming an inverting state.
Peaks in the BER trace correspond to these transitions, with nominal and inverted operation centered near $\num{280},\si{\micro\ampere}$ and $\num{305},\si{\micro\ampere}$, respectively.

Using these measurements, Figure~\ref{fig:figure4}b delineates the operating region boundaries extracted from the BER traces.
Boundaries were identified based on when the BER deviated within a $\pm\num{5}\si{\percent}$ range around 0.5, marking transitions between nominal and inverting states.
This plot shows that the operating margin depends on the write current for values less than $\num{30}\,\si{\micro\ampere}$, but becomes less dependent on write current for larger values.
While approximate, this plot provides a useful guide for selecting write and enable currents to ensure reliable operation of the memory cell.

Figure~\ref{fig:figure4}c explores the write-current dynamics at a fixed write temperature by performing linear sweeps of write-currents from $\num{0}\,\si{\micro\ampere}$ to $\num{300}\,\si{\micro\ampere}$, across multiple temperatures defined by varying enable currents for cell C3.
Additionally, we see that write currents beyond $\num{60}\,\si{\micro\ampere}$ lead to an increase in the minimum BER, indicating that a greater fraction of write pulses switch both hTrons to the normal state, leading to a higher probability of inverting the persistent current.

Extracting critical switching currents from these BER traces, as presented in Figure~\ref{fig:figure4}d, allowed identification of minimum and maximum operational write-currents.
Bounds are just estimates as measurement currents have some uncertainty.
This shows that increasing the temperature increases the margin between the minimum and maximum write-currents, but the maximum write-current is limited by the critical current of the right hTron.
Once the temperature exceeds the critical temperature, increasing the enable current does not increase the write current margin.

Figure~\ref{fig:figure4}e displays the minimum BER for each memory cell within the $\num{4}\times\num{4}$ array, achieved through systematic optimization of read and write voltages.
All cells were functional, with the fidelity (1-BER) in~\ref{fig:figure4}f showing that 15 out of 16 cells exhibit greater than 99.9\% fidelity, outperforming the previous state-of-the-art (dashed red line).
Cell C2 achieved 94\% fidelity, which may be affected by suppressed critical currents in a neighboring cell.
No systematic spatial variations were detected, underscoring the scalability and uniformity of the superconducting memory array operation.

Finally, Figure~\ref{fig:figure4}g examines memory retention, depicting BER as a function of delay between write and read operations.
The same BER measurement sequence was used.
The bias conditions were held constant to emphasize the temporal stability of the memory cell, rather than minimize BER at each time point.
Error bars corresponding to the standard deviation expected from binomial statistics with $\num{2e5}$ trials are smaller than the markers.
Extending the hold time to $\num{20}\,\si{\second}$ demonstrated robust performance, with zero errors across $\num{100}$ measurements.

Our achieved BER of $10^{-5}$ is comparable to other memory technologies, such as niobium kinetic inductance memories~\cite{ilin2021supercurrent}.
Lower BERs have been demonstrated in magnetic tunnel junction memories~\cite{nguyen2020cryogenic}, but consume greater energy per operation and have a larger footprint.
The demonstrated retention time of $20\,\si{\second}$ is limited only by the measurement duration, with no observed loss of the stored state, confirming the nonvolatile nature of the memory.

The calculated power and energy consumption for each memory operation are summarized in Table~\ref{tab:table1}, along with the measured pulse widths (at half maximum amplitude) and relative timing delays between input and enable pulses.
Write operations consume minimal energy ($\num{46}\,\si{\femto\joule}$) compared to enable pulses, which dominate the energy budget (up to $1256\,\si{\femto\joule}$ for EW). The read pulse energy is intermediate at approximately $\num{31}\,\si{\femto\joule}$.
Detailed calculations and methodologies for determining these parameters are provided in the supplementary information, and the pulse definitions correspond to those illustrated in Figure~\ref{fig:figure1}.

Table~\ref{tab:table1} lists the calculated power and energy values for each memory operation.
Additionally, we report the width of each pulse (at half max) and the relative delay between input and enable operations.
These values are limited both by the measurement setup and the device characteristics, and they were determined at a minimum bit error rate.
Faster operation and lower energy per switch has been demonstrated in single-cell memories~\cite{butters2021scalable,mccaughan2018kinetic,zhao2018compact}, and smaller hTron geometries are expected to further reduce these values~\cite{wang2025attojoule}.

The demonstrated performance metrics position SNM technology for integration into superconducting computing architectures where conventional CMOS memory is power-prohibitive.
This technology provides a solution for applications requiring ultra-compact, nonvolatile, zero-static-power memory, such as large-scale quantum error correction systems where memory density and power constraints are critical.

Several limitations constrain the current implementation and must be addressed for practical deployment.
The use of normal-metal enable lines, while providing precise thermal control, limits power efficiency and scaling potential due to resistive losses.
The destructive readout requires rewriting after each read operation, increasing access latency and energy consumption compared to non-destructive alternatives.
Temperature-dependent operation introduces thermal crosstalk considerations that may limit packing density without additional design constraints.
Finally, the demonstrated $\num{4}\times\num{4}$ array represents an initial proof-of-concept; scaling to kilobit-scale arrays will require addressing fabrication uniformity, yield optimization, and integration with peripheral addressing circuitry.

\begin{table}[ht]
    \caption{Measured and derived values for: write (W), read (R), enable-write (EW), and enable-read (ER) pulses. Pulse widths and delays were determined at 50\% of the peak amplitude. Values are approximate and based on the methodology described in the Methods section, taken from a single device operating at minimum bit error rate.}\label{tab:table1}
    \centering
    \begin{tabularx}{0.4\textwidth}{XX}
        \toprule
        $P_{W}$              & $1\,\si{\nano\watt}$      \\
        $P_{R}$              & $368\,\si{\nano\watt}$    \\
        $P_{EW}$             & $57.1\,\si{\micro\watt}$  \\
        $P_{ER}$             & $9.2\,\si{\micro\watt}$   \\
        \midrule
        $E_{W}$              & $46\,\si{\femto\joule}$   \\
        $E_{R}$              & $31\,\si{\femto\joule}$   \\
        $E_{EW}$             & $1256\,\si{\femto\joule}$ \\
        $E_{ER}$             & $202\,\si{\femto\joule}$  \\
        \midrule
        $\tau_{W}$           & $40\,\si{\nano\second}$   \\
        $\tau_{R}$           & $80\,\si{\nano\second}$   \\
        $\tau_{EW}$          & $22\,\si{\nano\second}$   \\
        $\tau_{ER}$          & $22\,\si{\nano\second}$   \\
        $\delta\tau_{W, EW}$ & $2\,\si{\nano\second}$    \\
        $\delta\tau_{R, ER}$ & $15\,\si{\nano\second}$   \\
        \bottomrule
    \end{tabularx}
\end{table}

\FloatBarrier{}
\section{Conclusion}\label{sec:Conclusion}

This work demonstrates a $\num{16}$-bit superconducting nanowire memory array with reliable read–write functionality at $\num{1.3}\,\si{\kelvin}$, achieving a minimum bit error rate of $10^{-5}$ at $\num{1}\,\si{\mega\hertz}$ with write and read energies of $\num{1.2}\,\si{\pico\joule}$ and $\num{0.2}\,\si{\pico\joule}$, respectively.
Memory states remain stable for at least $\num{20}\,\si{\second}$ without refresh, confirming the nonvolatile nature of the technology.

The architecture supports aggressive scaling through reduced cell footprint, interconnect pitch optimization, and advanced materials with higher kinetic inductance.
Several technical challenges remain: normal-metal enable lines limit power efficiency, destructive readout increases latency, and scaling to kilobit arrays will require addressing fabrication uniformity and peripheral circuitry integration.
Non-thermal switching approaches and operation at higher temperatures (up to $\num{4}\,\si{\kelvin}$ or beyond with high-$T_C$ materials) offer promising paths for improvement.

This work establishes superconducting nanowire memory as a compact, low-energy solution for cryogenic computing, providing a foundation for larger-scale memory systems essential to future superconducting and quantum computing architectures.

\section{Methods}\label{sec:Methods}

Memory cell operation was characterized using a combination of experimental measurements and circuit simulations.
The hTron thermal response was modeled using a behavioral framework~\cite{karam2024parameter} that relates the enable current to channel temperature, allowing prediction of switching thresholds and persistent current amplitudes.
Power and energy dissipation were calculated from measured pulse parameters, with write operations consuming $\num{46}\,\si{\femto\joule}$ and read operations $\num{31}\,\si{\femto\joule}$ (see Table~\ref{tab:table1}).

The $\num{16}$-bit memory arrays were fabricated using a three-step lithography process on thermal oxide wafers with $\num{23}\,\si{\nano\meter}$ NbN films ($T_c = \num{12.5}\,\si{\kelvin}$).
Measurements were performed at $\num{1.3}\,\si{\kelvin}$ using arbitrary waveform generators and room temperature amplifiers.
Bit error rates were characterized using sequences of write–read operations, with pulse amplitudes optimized via Bayesian optimization~\cite{head2020scikit} to minimize errors while maximizing operating margins.

The temperature of the hTron channel, $T_{\text{channel}}$, is computed from the enable current, $I_{\text{enable}}$, using a phenomenological heating model:

\begin{align}\label{eq:channel_temperature}
    T_{\text{channel}} = {\left[(T_{C}^4 - T_{\text{sub}}^4){\left(\frac{I_{\text{enable}}}{I_{\text{enable, max}}}\right)}^{\eta} + T_{\text{sub}}^4\right]}^{1/4}
\end{align}

Here, $T_{C}$ is the film's superconducting critical temperature, $T_{\text{sub}}$ is the substrate temperature, and $\eta$ is a dimensionless parameter that captures the thermal coupling between the nanowire and the substrate.
Using this temperature model, we compute the channel's critical current as:

\begin{align}\label{eq:critical_current}
    I_{C}(T_{\text{channel}}) = I_{C,0}{\left[1 - {\left(\frac{T_{\text{channel}}}{T_{C}}\right)}^{3}\right]}^{2.1}
\end{align}

Detailed methods including thermal modeling equations, fabrication procedures, experimental setup specifications, and power calculations are provided in the Supplementary Information.
\backmatter{}
\section*{Acknowledgments}
The authors thank Evan Golden and Phillip D. Keathley for their review during the preparation of this manuscript.
This work was funded by the DOE Office of Science Research Program for Microelectronics Codesign through the project `Hybrid Cryogenic Detector Architectures for Sensing and Edge Computing Enabled by New Fabrication Processes' (LAB 21\-2491).
Fabrication was carried out in part through the use of MIT.nano's facilities, with technical guidance from James Daley and Mark Mondol.
Initial fabrication development was funded through the Breakthrough Starshot Foundation.
O. Medeiros acknowledges support through the National Defense Science and Engineering Graduate (NDSEG) Fellowship Program.
A. Simon acknowledges NSF GRFP and MIT Vanu Bose Presidential fellowships.
R. Foster acknowledges the Alan McWhorter Fellowship.

\end{document}